\documentstyle[aps,prl,twocolumn,epsfig]{revtex}
\newcommand{\nb}{\nolinebreak}
\begin{document}
\draft
\wideabs{
%\title{Surface adsorption on carbon nanotubes as a physical realization of
%quantum spin tubes: geometric frustration and magnetization plateaus}
%\title{Adsorption on carbon nanotubes: quantum spin tubes,
%geometric frustration and magnetization plateaus}
\title{Adsorption on carbon nanotubes: quantum spin tubes, magnetization
plateaus, and conformal symmetry}
\author{Dmitry Green$^a$ and Claudio Chamon$^b$}
\address{$^a$ Department of Physics, Yale University,
New Haven, CT 06520\\ $^b$ Department of Physics, Boston University,
Boston, MA 02215}

\maketitle
%\twocolumn

\begin{abstract}
  We formulate the problem of adsorption onto the surface of a carbon
  nanotube as a lattice gas on a triangular lattice wrapped around a
  cylinder.  This model is equivalent to an XXZ Heisenberg quantum
  spin tube.  The geometric frustration due to wrapping leads
  generically to four magnetization plateaus, in contrast to the two
  on a flat graphite sheet.  We obtain analytical and numerical
  results for the magnetizations and transition fields for armchair,
  zig-zag and chiral nanotubes.  The zig-zags are exceptional in that
  one of the plateaus has extensive zero temperature entropy in the
  classical limit.  Quantum effects lift up the degeneracy, leaving
  gapless excitations which are described by a $c=1$ conformal field
  theory with compactification radius quantized by the tube
  circumference.
\end{abstract}
}

Monolayer adsorption of noble gases onto graphite sheets has proven to
be an interesting problem both theoretically and
experimentally\cite{Bretz,Schick,Murthy}.  Many of the observed
features can be understood within a lattice gas model, where the
underlying hexagonal substrate layer forms a triangular lattice of
preferred adsorption sites.  An equivalent formulation is in the
language of spin models on a triangular lattice, where the repulsion
between adsorbed atoms in neighboring sites translates into an
antiferromagnetic Ising coupling.  The frustration of the couplings by
the triangular lattice leads to the rich phase diagram of the
monolayer adsorption problem\cite{Schick}. Introducing hopping adds
quantum fluctuations, further enriching the phase
diagram\cite{Murthy,Sondhi}.

In this paper we address what happens if, in addition to the
triangular lattice frustration, one has an extra geometric frustration
due to periodic boundary conditions.  In fact, such a system is
physically realized by a single walled carbon
nanotube\cite{Dresselhaus}, which may be viewed as a rolled graphite
sheet.  In this context, adsorption has been the subject of growing
experimental and theoretical interest \cite{Stan,Cole} spurred by
potential applications. Stan and Cole\cite{Stan} have considered the
limit of non-interacting adatoms at low density, finding that they are
localized radially near a nanotube's surface at a distance comparable
to that in flat graphite ($\sim 3{\mbox \AA}$). In that work, it was
sufficient to omit the hexagonal structure of the substrate.  However,
the corrugation potential selects the hexagon centers as additional
commensurate localization points\cite{Bretz}. In view of the
similarity to flat graphite, we include both the substrate lattice and
adatom interactions and consider a wider range of densities. In fact,
very recently, it has been shown \cite{ColeNew} that the adsorbate
stays within a cylindrical shell for fillings less than $\approx 0.1
/{\mbox \AA}^2$ (or $\approx 0.5$ adatom/hexagon), justifying the
densities studied here.

\begin{figure}
\noindent
\center
\vspace{-.5 in}
%\hspace{.9 in}
\epsfxsize=2.0 in
\epsfbox{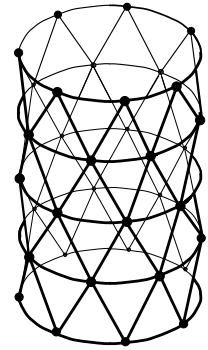}
\vspace{-.3 in}
\caption{Adsorption sites on a $(7,0)$ zig-zag nanotube}
\label{Tube}
\end{figure}

The adsorption sites thus form a triangular lattice wrapped around a
cylinder.  The geometrically inequivalent ways in which the wrapping
is realized are labeled by two integers, $(N,M)$; the case $(N,N)$ is
known as the armchair tube, $(N,0)$ is the zig-zag, and all others are
chiral.  The geometric frustration is present whenever the wrapping
destroys the tripartite nature of the triangular lattice, which occurs
when $(N-M) {\rm mod}\;3$ is non-zero (this criterion is familiar in
the context of electronic conductivity\cite{Dresselhaus}).  Thus, both
zig-zag and chiral tubes can be frustrated geometrically, whereas
armchair tubes cannot.  In Fig.~\ref{Tube} we show the tube obtained
from the $(7,0)$ zig-zag.  In terms of the equivalent spin systems,
this is similar to recent models of ``spin tubes''\cite{Andrei}.
When the adsorbed gas is a hard-core boson, the lattice gas is defined
by the Bose-Hubbard Hamiltonian\cite{Murthy,Auerbach}
\begin{equation}
 \label{eq:Hubbard}
{\cal H}\! =\! -t\sum_{\langle ij\rangle} b_i^\dagger b_j + b_j^\dagger
 b_i
 + V\sum_{\langle ij\rangle} n_i n_j - \mu\sum_i n_i~,
\end{equation}
where $n_i$ is the boson density at site $i$, $V$ is the nearest neighbor
repulsion and $t$ is the hopping amplitude. In the equivalent Heisenberg spin
representation,
\begin{equation}
 \label{eq:Spin}
{\cal H}\! =\!  -2t  \sum_{\langle ij\rangle} S^x_i S^x_j\! +\! S^y_i S^y_j
 + V \sum_{\langle ij\rangle} S^z_i S^z_j \!-\! H\sum_i S^z_i ~,
\end{equation}
where $S^z_i=n_i-1/2$ and $H=\mu-3V$ is an effective external magnetic
field. Throughout the paper we will use the spin and density
representations interchangeably.

The Ising limit, in which hopping is not allowed, already contains
many interesting features.  We start the analysis in this regime,
obtaining the phase diagram as a function of the magnetic field, and
then consider quantum fluctuations perturbatively in $t/V$.  We
summarize our results first.

\begin{figure}
\noindent
\center
\epsfxsize=2.5in 
\epsfbox{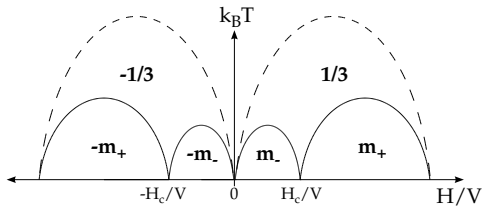}
\caption{Phase diagram of the $(N,M)$ tube}
\label{PhaseDiagram}
\end{figure}

The phase diagram in the temperature-magnetic field plane of a typical
tube is shown in Fig.~\ref{PhaseDiagram}.
When the index $q=(N-M){\rm mod}\;3$ is $1$ or $2$, we find four lobes
(solid lines), corresponding to two plateaus with magnetizations $m_-
< 1/3$ and $m_+ > 1/3$.  Here, we use the standard Ising notation in which
spin is $\pm1$.  Note that the plateaus are real phases only at
zero temperature because the tube is one-dimensional.  At finite
temperature, the boundaries should be interpreted as crossovers.
Nonetheless, deep within a lobe, at $k_BT\ll V$, the magnetizations
are well-defined.  Specifically, for $q=1$, we obtain the exact
expressions
\begin{eqnarray}
&\;& m_+=\frac{1}{3}\left(1+\frac{2}{2M+N}\right) \quad
m_-=\frac{1}{3}\left(1-\frac{2}{2N+M}\right) \nonumber\\
&\;&\hspace{2cm}
H_c=\left(4-\frac{2M}{N+M}\right)V
\label{eq:Exact}
\end{eqnarray}
The complementary case of $q=2$ is obtained by interchanging
$N\leftrightarrow M$.  On the other hand, those tubes without
geometric frustration ($q=0$) behave similarly to the flat sheet
(dotted lines) which has only two lobes with magnetizations $\pm 1/3$
\cite{Schick}.  As the tube perimeter approaches the flat sheet limit, one
expects that the geometric frustration becomes irrelevant.  Indeed, as
$N$ or $M\rightarrow\infty$, $m_+$ and $m_-$ squeeze $1/3$ as the
inverse of the tube diameter and become indistinguishable.  Beyond the
lobes, where the field is strong enough to overcome all nearest
neighbor bonds ($|H|/V>6$ at $k_BT=0$), the tube is fully
polarized.  The filling fractions are
obtained from the magnetizations by $m=-2(n-1/2)$.  The phase diagram,
however, is more easily visualized in terms of spin since
spin reversal, $m\leftrightarrow-m$, corresponds to particle-hole
symmetry, $n\leftrightarrow1-n$.

\begin{figure}
\noindent
\center
\vspace{-0.1 in}
\epsfxsize=3.2 in
\epsfbox{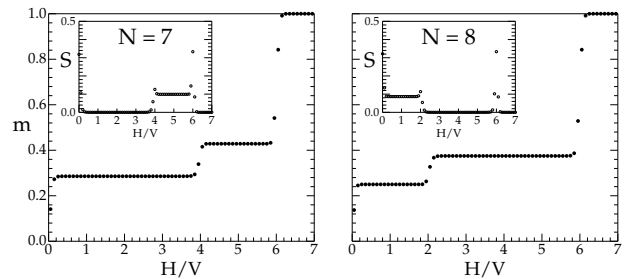}
%\vspace{.3cm}
\caption{Magnetization and entropy per site at $k_BT=0.05V$}
\label{Magnetization}
\end{figure}

We have verified this prediction numerically by transfer matrix
methods for zig-zag tubes up to $N=11$ and for the chiral tubes up to
$N+M=7$.  Although $7$ is probably too small to be physical, we
believe that the arguments in this paper generalize to any tube.  In
Fig.~\ref{Magnetization} we display sample data for two zig-zag tubes
with different $q$: $(7,0)$ and $(8,0)$.  The magnetization curves
show clear plateaus whose values and transition fields match those
predicted by Eq.~(\ref{eq:Exact}).  By increasing the temperature and
following the evolution of the plateaus, we generate the phase diagram
above.

\begin{figure}
\noindent
\center
\vspace{-1.0 in}
\epsfxsize=2.5 in
\epsfbox{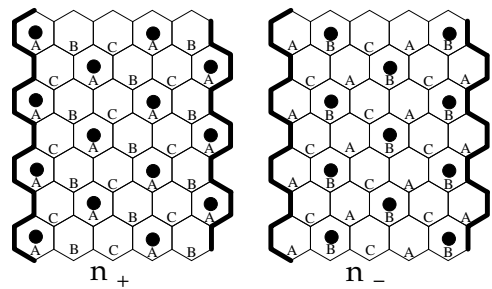}
\vspace{-.6 in}
\caption{Fillings and zipper of the $(5,0)$ zig-zag tube. $n_\pm$ corresponds to $m_\mp$}
\label{TubeFillReg}
\end{figure}

We find that a rather interesting feature of the zig-zag $(N,0)$ tubes
emerges, making them exceptional.  The insets in
Fig.~\ref{Magnetization} indicate an extensive entropy at zero
temperature, which has plateaus, too.  Upon enumerating the degenerate
space explicitly, we shall show that the entropy is exactly $s=({\rm
ln}2)/N$ and that it occurs in $m_+$ for $q=1$ and in $m_-$ for $q=2$.
In the presence of hopping, the non-degenerate plateaus retain their
gaps, whereas the degenerate ones become correlated states with a
unique ground state and {\it gapless} excitations.  More precisely,
conformal invariance develops and the effective theory has central
charge $c=1$ with a compactification radius, $R$, quantized by the
tube circumference, $R=N$.
In order to understand the magnetizations and nature of the geometric
frustration, it is more intuitive to use the original bosonic picture.
As a result of hard-core repulsion on the infinite graphite sheet, the
$m=1/3$ plateau corresponds to filling one of the three sublattices,
$A, B$ or $C$, of the triangular lattice.  This configuration
minimizes the repulsion, $Vn_i n_j$, while maximizing the filling, $\mu
n$. It is natural to try the same for nanotubes, as we illustrate in
Fig.~\ref{TubeFillReg} for $(5,0)$.
Upon wrapping, however, the thick vertical lines are identified and
the lattice is no longer tripartite.  In fact, the number of
sublattice sites is no longer equal, and there is a mismatch along the
thick line, which we term the ``zipper''.  On the left we fill the $A$
sublattice, obtaining the filling fraction $n_+=2/5$, and on the right
either $B$ or $C$ may be filled with the result that $n_-=3/10$.  For
general $(N,0)$ there are $2N$ hexagons in the unit cell, and the
filling fractions are $n_+=\lceil 2N/3\rceil/2N$ and $n_-=\lfloor
2N/3\rfloor/2N$, where $\lceil x\rceil$ and $\lfloor x\rfloor$ denote
the larger and smaller of the two bounding integers of $x$,
respectively.  The magnetizations in Eq.~(\ref{eq:Exact}) follow
directly by using the correspondence $m=-2(n-1/2)$.  Furthermore, due
to the sublattice mismatch, the number density of adjacent particles,
$n_b$, may be non-zero.  In the case of $(5,0)$, there are two broken
bonds per unit cell in $n_+$, and none in $n_-$.  This result
generalizes to any $q=2$ zig-zag tube: $n_{b+}=2/2N$ and $n_{b-}=0$.
For $q=1$, the argument goes through as before, except that
$n_{b+}=1/2N$.  We summarize this compactly by $n_{b+}=q/2N$.

Substituting these fillings into the Hamiltonian (\ref{eq:Hubbard})
yields two energies per site, $e_{\pm}(\mu)=Vn_{b\pm}\!-\!\mu
n_{\pm}$.  The transition occurs when these levels cross: $e_+=e_-$,
or
\begin{equation}
\frac{q}{2N}-\mu\frac{\lceil 2N/3\rceil}{2N}=-\mu\frac{\lfloor
2N/3\rfloor}{2N}
\label{eq:Transition}
\end{equation}
Solving for $\mu$ and using the correspondence
$H=\mu-3V$ gives precisely the critical field in Eq. (\ref{eq:Exact}).
In particular, this explains why there are exactly two independent
plateaus.  Note that, for the special case of the zig-zags, the
critical field depends only on $q$ and not on $N$ per se.

In the above analysis, we have made only one assumption, namely that
the zipper runs parallel to the tube axis. In general, the zipper may
wind helically around the tube or wiggle sideways. However, in all the
cases that we considered, the straight zipper has the lowest energy, and moreover, our transfer matrix
computations, which are blind to this assumption, are consistent with
our analysis.

The chiral tubes are different. Due to their geometry the zipper is
forced to wind, but, again, we find that the choice of the straightest
possible zipper reproduces our numerics for $N+M$ up to
$7$.  The determination of the fillings and level crossings is much
more involved than that of the zig-zag, and we leave it for a more
detailed paper.  In any case, our analysis reveals that the plateaus
in a chiral tube are not macroscopically degenerate, so that the
zig-zags are at a special degenerate point.

Having understood in detail the Ising limit, we now turn on a small
hopping, $t\ll V$, that introduces quantum fluctuations.  Deep within
a plateau, the substrate is maximally filled since adding a particle
increases $n_b$.  Consequently, all plateaus begin with a classical
gap of order $V$, and we work in the Hilbert space of the
classical ground states. Those plateaus which have only a
discrete symmetry must retain their gaps, but the macroscopically
degenerate plateaus are more complicated.

\begin{figure}
\vspace{-1.5 in}
\center
\epsfxsize=3.3 in
\noindent
\epsfbox{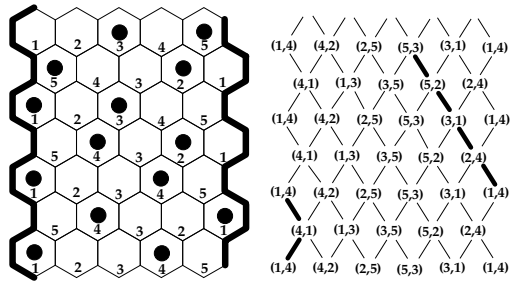}
\vspace{-1.3 in}
\caption{LEFT: Typical configuration in $n_+$ (or $m_-$) of the
$(5,0)$ tube.  Alternating numbering within layers allows a symmetric
description from bottom-to-top or top-to-bottom.  RIGHT: Allowed
states as paths on a wrapped square lattice.  The vertex labels may be
dropped.}
\label{Hop-Lattice}
\end{figure}

Let us reconsider the $n_+$ filling of the $(5,0)$ tube in
Fig.~\ref{TubeFillReg}.  Notice that a particle may hop laterally by
one site without changing $n_b$, as we illustrate in
Fig.~\ref{Hop-Lattice}, left.  Imagine building a typical $n_+$ state
layer-by-layer from top to bottom, with a total of $L$ layers.  Each
new layer must add exactly two filled sites and one nearest-neighbor
bond ($n_b=1/5$).  This constraint implies that no two adjacent sites
may be occupied within a layer; if they were, then, to conserve $n_b$,
two adjacent sites must be occupied in the next, and so on up the
tube.  However, this state is not connected to any other by a single
hop.  Similarly, the particles cannot hop from layer to layer because
this adds another intra-layer bond.
An allowed state can be represented as a string of occupied sites,
$\{\sigma_i\}$, $i=1,\ldots,L$, which in our example is
$\{\cdots(5,3)(5,2)(3,1)(2,4)(1,4)(1,4)\cdots\}$.  At each layer,
there are exactly two possibilities for the following one.  For
example, $(1,4)$ can be followed by $(1,4)$ or by $(2,4)$.  However,
the total number of possibilities at any given level is five.
Fig.~\ref{Hop-Lattice} (right) summarizes this structure succinctly as
a {\em square} lattice wrapped on the cylinder.  A typical state,
then, is a lattice path along the tube.

Generalizing to $(N,0)$, we find $N$ possible states in each layer and
two in the succeeding one, and the structure of states is again that
of a wrapped square lattice with $N$ squares along the circumference.
The dimension of the Hilbert space is the number of lattice paths,
$N2^L$, so that in an infinitely long tube, the entropy per site is
exactly $({\rm ln}2)/N$, as claimed earlier.  Notice that constrained
paths introduce correlations along the length of the tube, despite the
absence of inter-layer hopping.

The matrix elements of the projected Hamiltonian connect only those
states that differ by a single hop:
\begin{equation}
\left<\{\tau\}|{\cal H}|\{\sigma\}\right> = 
\Big\{\begin{array}{c}
-2t\quad{\rm if}\quad\sum_i\delta_{\sigma_i \tau_i} = L-1\\
0\quad{\rm otherwise}
\end{array}
\label{eq:Hopping}
\end{equation}
We diagonalize this Hamiltonian numerically with periodic boundary
conditions for system sizes up to $N=11$ and $L=10$.  Additionally, we
can obtain the ground state energy up to $L=16$ due to the sparseness
of ${\cal H}$.  We will fix $2t=1$ in what follows. 

\begin{figure}
\noindent
%\hspace{2 in}
\center
%\hspace{.8 in}
\epsfxsize=3.0in
\epsfbox{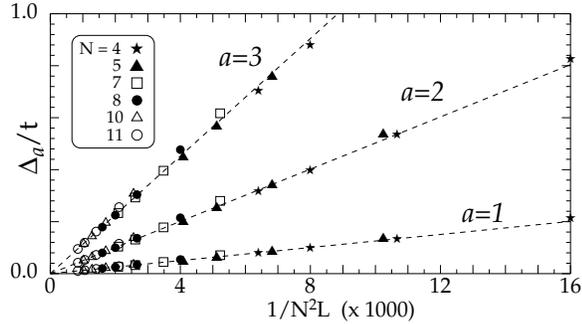}
\vspace{.2cm}
\caption{The gap scales as $1/N^2L$.}
\label{Gap}
\end{figure}

We find that the degeneracy is lifted and the ground state becomes
unique and uniform.  The ground state energy, $E_0(L)$, follows
$E_0\!\sim\!-0.607L\!-\!\pi c/6$ with $c\nb\!\sim\nb\!1.005$.  The lowest
$N\!-\!1$ excited states are given by $\Delta_a=a^2\Delta/(N^2L)$,
with $\Delta=12.9\pm 0.5$, which is shown in Fig.~\ref{Gap} for
$a=1,2,3$.  All of these levels are doubly degenerate.
This ground state energy and spectrum are in perfect agreement with a
conformally invariant bosonic theory with central charge $c\!=\!1$
compactified on a radius $R\nb\!=\nb\!\zeta N$. We take the Lagrangian
density ${\cal L}\nb =\nb\frac{1}{8\pi}[v^{-1}(\partial_t\phi)^2\nb
-\nb v(\partial_x\phi)^2]$, where $v$ is a velocity, and the
compactification is defined by $\phi\equiv\phi+2\pi R$.  The zero mode
energies of this theory are\cite{CFT}
\begin{equation}
E^0_{a,b}=\frac{2\pi v}{L} \left(\frac{a^2}{R^2}+\frac{b^2 R^2}{4}\right) ,
\label{ZeroModes}
\end{equation}  
where $a,b$ are integers that label topological momenta and windings
of $\phi$.  Right- and left-moving oscillator modes of energy
$\omega_n=vk_n$, where $k_n=2\pi n/L$, also appear in the spectrum,
but for $a<N$ the zero modes are the lowest.  Our spectrum in
Fig.~\ref{Gap} corresponds to $E^0_{a,b}$ with $b=0$.  To fix $\zeta$,
we look at higher low-lying levels (which also scale like $1/L$). We
find that the $N$'th excitation energy is independent of $N$ and
quadruply degenerate.  This can happen only if the $N$'th zero mode,
$E^0_{\pm N,0}=2\pi v/\zeta^2L$, is degenerate with the lowest
oscillator mode, $\omega_{\pm 1}=2\pi v/L$, which fixes $\zeta=1$.
Thus, the compactification radius is $R=N$.  The velocity can be read
off from the slopes in Fig.~\ref{Gap} as $v=\Delta/2\pi$.  The rest of
our spectrum is consistent with these parameters.
%For instance, we find a
%unique, zero-momentum state with $a=b=0$, which consists of one right-
%and one left-moving oscillator mode with $n=2\pi/L$ at energy $E=2E_1$.

One observable consequence of conformal symmetry is that the low
temperature heat capacity is fixed by $c$\cite{CFT}:
\begin{equation}
C=c\frac{\pi k_B^2}{3}T=\frac{\pi k_B^2}{3}T
\nonumber
\end{equation}

It is noteworthy that, even though the dispersion of the oscillator
modes is independent of $N$, the spectrum remembers, via the
zero-modes, the finite radius of the nanotube. Furthermore, $R$ is
quantized by $N$; in the language of Luttinger liquids, this means
that the Luttinger parameter is fixed by topology, similarly to the
case of edge states in a fractional quantum Hall fluid\cite{Wen}, and
in contrast to quantum wires (where the Luttinger parameter can vary
continuously). Because there is no inter-layer hopping, $\phi$ is tied
to transverse, rather than to longitudinal, density fluctuations along
the tube.  We will present a detailed analytical derivation of the
effective theory from the lattice in Fig.~\ref{Hop-Lattice} elsewhere.

Let us briefly view the spin tube as a quantum spin ladder to see if
it yields the zero gap.  A standard approach is to use a
Lieb-Schultz-Mattis (LSM) argument, in which the spins are deformed
slowly along the length\cite{Affleck}.  Applying it to our tube, we
find that a plateau is gapless if $S-M$ {\it is not} an integer, where
$S$ is the total spin and $M$ the magnetization per layer.  Using
$S=N/2$ and the magnetizations from Eqn.~(\ref{eq:Exact}), we find
that $S-M$ {\it is} an integer in the macroscopically degenerate
plateaus, so that the LSM argument is insufficient in this case. A
conclusive argument must take the geometric frustration into account.

Before concluding, we should point out that the geometry of the
$(2,0)$ tube is special; all sites in adjacent layers are
interconnected.  As a result, all of its plateaus have an extensive
entropy, and we find that hopping opens a gap in both plateaus.  In
fact, this tube can be written as a spin chain that has been studied
at isotropic coupling\cite{Sakai}, $-2t=V$.  Two plateaus were found in
this case, and it is tempting to speculate whether the two regimes are
connected adiabatically.

In conclusion, we have studied the problem of monolayer adsorption on
carbon nanotubes and identified several interesting filling fraction
plateaus. Since the difference between the plateaus decreases slowly,
as the inverse of the tube diameter, experimental measurement should
be feasible for large enough tubes. We have identified the zig-zag
tubes as exceptional, in which the geometric frustration together with
quantum fluctuations lead to conformal symmetry. This system is a
physical realization of quantum spin tubes.

The authors wish to thank C.~Buragohain, M.~El-Batanouny, E.~Fradkin,
N.~Read, C.~Nayak, M.~Vojta, and X.-G.~Wen for helpful comments.
Support was provided by the NSF Grant DMR-98-18259(D.~G.),
DMR-98-76208 and the Alfred P. Sloan Foundation (C.~C.).

%By making a slow rotation of spins along
%a tube with periodic boundary conditions, it is determined that a
%necessary, {\it but not sufficient}, condition for the existence of a
%gap is for $Q(S-m)$ to be an integer [Affleck], where $Q$ is the
%periodicity of the ground state, $S$ is the total spin per layer, and
%$m$ is the magnetization.  In our case, since there is a classical
%gap, we find that the only way for it to persist in the quantum regime
%is for $Q$ to be an even integer, so charge density wave order
%survives.  

%
%There are many tantalizing questions raised by our
%investigations.  For instance, what is the interplay between the
%nanotube conductivity, an interesting phenomenon in its own right, and
%adsorption; can carbon-carbon bond distortion due to the adsorbed gas
%modify the electronic conductivity?  There are undoubtedly many
%surprises left in the details of the transitions and cross-over
%regions of our phase diagram.


\vspace{-0.3 in}
\begin{references}
\vspace{-0.6 in}
\bibitem{Bretz} M.~Bretz, J.~Phys.~Col. {\bf 39}, C6/1348-51 (1978).
\bibitem{Schick} M.~Schick, J.~S.~Walker and M.~Wortis, \prb {\bf 16}, 2205 (1977).
\bibitem{Murthy} G.~Murthy, D.~Arovas, A.~Auerbach, \prb {\bf 55}, 3104 (1997).
\bibitem{Sondhi} R.~Moessner, S.~L.~Sondhi and P.~Chandra, cond-mat/9910499.
\bibitem{Dresselhaus} {\it Science of Fullerenes and Carbon Nanotubes}, M.~S.~Dresselhaus, G.~Dresselhaus and P.~C.~Eklund, Academic Press (1996).
\bibitem{Stan} G.~Stan and M.~W.~Cole, Surf. Sci. {\bf 395}, 280 (1998).
\bibitem{Cole} G.~Stan, M.~J.~Bojan, S.~Curtarolo, S.~M.~Gatica, M.~W.~Cole, cond-mat/0001334.
\bibitem{ColeNew} S.~M.~Garcia, G.~Stan, M.~M.~Calbi, J.~K.~Johnson and M.~W.~Cole, cond-mat/0004229.
\bibitem{Andrei} E.~Orignac, R.~Citro and N.~Andrei, cond-mat/9912200.
\bibitem{Auerbach} {\it Interacting Electrons and Quantum Magnetism}, A.~Auerbach, Springer-Verlag (1994).
\bibitem{CFT} P.~Ginsparg in {\it Fields, Strings and Critical Phenomena}, E.~Brezin and J.~Zinn-Justin, eds., Elsevier (1988); J.~L.~Cardy, {\it ibid}.
\bibitem{Wen} X.~G.~Wen, \prb {\bf 41}, 12838 (1990).
\bibitem{Affleck} M.~Oshikawa, M.~Yamanaka and I.~Affleck, \prl {\bf 78}, 1984 (1997).
\bibitem{Sakai} T.~Sakai and N.~Okazaki, cond-mat/0002113.
\end{references}
\end{document}